\documentclass[manuscript]{aastex}

\usepackage{psfig}

\slugcomment{Accepted in {\it Astrophysical Journal Letters}}
\shortauthors{Howell \& Ciardi}
\shorttitle{LL And \& EF Eri}

\begin{document}

\title{Spectroscopic Discovery of Brown Dwarf-like Secondary Stars \\ 
in the Cataclysmic Variables
LL Andromedae and EF Eridani}

\author{Steve B. Howell} 

\affil{Astrophysics Group, Planetary Science Institute, \\
Tucson, AZ  85705,
howell@psi.edu}

\and

\author{David R. Ciardi}

\affil{Department of Astronomy, University of Florida, 
Gainesville, FL 32611, \\
ciardi@astro.ufl.edu}

\begin{abstract}
Infrared spectroscopic observations of LL Andromedae and EF Eridani
are presented.
Our K band spectrum of LL Andromedae reveals 
the presence of methane absorption in the secondary star 
indicating an effective temperature $\la$ 1300K, 
similar to a ``T" type methane brown dwarf. 
The secondary star in EF Eridani is seen to be 
warmer with an effective 
temperature of $\sim$1650K and a spectroscopic classification 
consistent with an L4-5 star.
Both secondaries have theoretical mass estimates of 0.04-0.055 M$_{\odot}$ (40-55
M$_{\rm{Jupiter}}$).
Our IR spectroscopic observations of LL And and EF Eri provide the
first direct proof of the theoretical prediction that post-period minimum
cataclysmic variables contain brown dwarf-like secondary stars.
\end{abstract}


\newpage

\section{Introduction}

Theoretical work on the evolution of cataclysmic variables (Howell, Rappaport, and
Politano 1997; Howell, Nelson, Rappaport 2001)
indicates 
that the majority of the
present-day population will have very short orbital periods and
should contain low mass ((M$_{\rm{sec}}$ $\la$ 0.06 $M{_\odot}$), 
brown dwarf-like secondary stars.
An observationally defined group, the 
Tremendous Outburst Amplitude Dwarf novae or TOADs (Howell, 
Szkody \&
Cannizzo 1995; Politano, Howell, \& Rappaport 1998) 
has been speculated as being examples of
this type of cataclysmic variable (CV).
The orbital period of CVs evolve to shorter period until a
period minimum of about 78
minutes is reached. At this time, the binary evolves to 
{\it longer} orbital periods with
the secondary star continuing to lose mass and 
becoming a brown dwarf-like object. The secondaries are best described
as ``brown dwarf-like" as most
started their lives as normal, lower main
sequence (G-M) dwarfs. 

LL And has an orbital period of near 78 minutes, based on 
superoutburst observations
by Howell \& Hurst (1994).
Recently, Szkody et al., (2000) presented a minimum light 
(V=19.9) optical spectrum (4000-7000\AA) of LL And obtained 3 years past
its last superoutburst.
Their spectrum revealed weak, 
double-peaked line emission and a rising blue 
continuum which they fit with an 11,000K, 0.6M${_\odot}$ 
white dwarf. They found no evidence for the secondary star
and 
concluded that the weak
line emission and essentially absent disk contribution were indications 
of an extremely low
mass transfer rate; the system being in a 
``deep quiescence" given the 3 year time passage since 
LL And's last superoutburst.

EF Eri (P$_{orb}$=81 min) contains a white dwarf with a strong magnetic field, 
B=18 MG, as determined 
from 
high state IR spectral observations made in 1992 (Ferrario et al. 1996).
At that time, the K band spectral appearance of EF Eri was typical of a high
state polar, strong H and He emission and broad continuum modulation due to
cyclotron humps with no indication of the secondary star.

EF Eri has a V magnitude of 14.5 during high states but was at V=18 in
February 1997 (Wheatly \& Ramsey 1998). 
The low state spectrum obtained by Wheatly \& Ramsey revealed only weak
line and cyclotron emission.
According to AAVSO records, EF Eri has been fainter than
V=17.5 for almost four years now, remaining in an extended low state.
Beuermann et al. (2000) modeled the spectrum
obtained by Wheatly \& Ramsey and fit it with a cool WD (T$_{eff}$=9500K, the
coolest known WD in a CV).
These authors set strict observational limits that the secondary star must
be later then M9V, have T$_{eff}\la$ 2000K, and probably a mass $\la$
0.06 M$_{\odot}$.

To date, there have been a few papers providing indirect 
observational evidence that
some short period CVs contain sub-stellar secondaries 
(eg., AL Com - Howell et al., 1998; WZ Sge - Ciardi et al., 1998; V592 Her
- van Teeseling et al., 1999; EF Eri - Beuermann et al., 2000). 
However, until now, 
no direct unambiguous observational 
proof has been presented. 
This {\it letter} reports the first direct observational evidence 
for a brown dwarf-like secondary star in a 
TOAD and provides the first spectroscopic discovery
of brown dwarf-like secondary stars
in the CVs LL And and EF Eri.

\section{Observations}

K-band spectroscopy of LL And was obtained on UT 2000 October 20 and 
of EF Eri on UT 2000 December 2 using
the 3.8~m United
Kingdom Infrared
Telescope (UKIRT). 
The observations covered the wavelength range from 
$1.85-2.47$\micron~and were made with 
the UKIRT Cooled Grating
Spectrometer (CGS4) equipped with a 256 $\times$ 256 InSb
array. 
Details of the data reduction
procedures can be found in Howell et al., (2000).

Our 1 hr K band observation of LL And (Fig. 1) is
seven years past superoutburst
and shows no evidence for continuum accretion or line emission,
only the secondary star is seen. It appears that all mass transfer has ceased
in LL And.
Convolution with a K filter yields
K$\sim$17.5-18 mag for LL And.
Our 2 hr K band spectrum of EF Eri (Fig. 1) shows no evidence of line emission 
or continuum accretion
flux indicating that mass transfer is stopped.
Convolution with a K filter yields
K=14.6$\pm$0.2 mag for EF Eri. 
We note here that in these systems, the
WD will contribute $\la$4\% of the flux at 2.2 microns (see Ciardi et al., 1998).
Both secondary stars are clearly very different from 
any previously obtained IR spectra
of any CV (See Dhillon et al., 2000 and refs therein).

\section{Discussion}

Our spectrum of the secondary star in LL And contains the
ubiquitous blue-ward broad water absorption band common in many cool red objects,
possible CO absorption, and
shows a continuum break at 2.2 microns.
The spectrum of LL And has a similar appearance
to the bona fide methane ``T" type brown dwarf SDSS 1254 
(Leggett et al., 2000b and references therein). Given this similarity, the strength
of the water feature, and the expected cool temperature of the secondary, we
attribute the 2.2 micron break in LL And to methane; a common feature in
very cool, low mass objects (Leggett et al., 2000b).
The presence of CH$_4$ sets
an upper limit on the effective temperature of the secondary star in LL And
of $\sim$1300K (Leggett et
al., 2000a).
The yet unidentified broad features seen throughout the
spectrum of LL And are similar to equally unidentified features seen in an optical
spectrum of the faint TOAD AL Com (Howell et al., 1998). These absorption
features could be caused
by dust formation in the cool secondary atmosphere 
based on the similarity of these broad absorptions with dust features seen in 
model spectra of cool dusty brown dwarf photospheres (Hauschildt 2000).
 
EF Eri shows the typical broad water absorption bands seen in cool
stars (Leggett et al., 2001) with possible CO absorption 
at 2.3 microns and red-ward. 
The broad blue-ward water feature provides continuum depth and slope information
which can be used to place
constraints on the effective temperature 
of cool stars. Defining an index for the water band, 
D$_{H_2O}$=$\frac{F_{2.265\mu}}{F_{2.0\mu}}$,
we find (Figure 2) that a good relation exists for stars of type M0 to L9,
albeit not single valued. D$_{H_2O}$ breaks down for the T stars due to the
presence of CH$_4$ absorption.
For EF Eri, D$_{H_2O}$=1.29 indicating either spectral type M6 or L4-5
and the lack of methane absorption
at 2.2 microns sets a lower limit of T$_{eff}$$\sim$1300K.
An M6V secondary 
star will not fit in the Roche lobe of EF Eri given its 
short orbital period. This fact, along with
the ``later than M9" constraint imposed by 
Beuermann et al. (2000), lead us to conclude that the secondary star in EF Eri
is a brown
dwarf-like object of spectral type near L4-5, with T$_{eff}$$\sim$1650K.
However, we note that the 2.3 micron continuum break due to CO absorption
is not as strong 
as in the single stars GD 165B (L4) and DNS 1228AB (L5)
(T$_{eff}$ = 1750-1550K, Leggett et al., 2001).

The evolutionary state of LL And and EF Eri can be considered by using
the CV secular evolution code described in Howell et al., (2001).
The secondary star properties for LL And and EF Eri
can be modeled by
considering an approximate orbital period for both of 80 minutes and 
examining both {\it pre}- and {\it post}-orbital period minimum cases.
If we assume that either
system is a short orbital period CV which 
has not yet reached the
orbital period minimum, its secondary star would have a mass 
near 0.11M${_\odot}$ and an
effective temperature of $\sim$3100K, yielding an approximate spectral type 
of M5V.
Our IR spectral evidence disallows this possibility and favors 
LL And and EF Eri as being very old CVs containing cold, low mass 
brown dwarf-like secondary stars 
with masses near 0.04M$_{\odot}$ (42 M$_{\rm{Jupiter}}$), radii near 0.1
R$_{\odot}$, and effective temperatures of 1300K and 1650K respectively. 
However, this mass estimate depends on the exact value of the orbital period minimum
as calculated by HNR. An alternative method, fairly independent of the exact
value or even the existence of the period minimum, is to use a theoretical
evolutionary calculation of T$_{eff}$ vs. M$_2$ and estimate the secondary mass
from its effective temperature based on our spectra.
Using the effective temperatures determined above, both secondary masses would be
near 0.055 M${_\odot}$.

Assuming that the absolute luminosity 
of the secondary star in LL And 
is a close approximation to that of a field methane 
brown dwarf (M$_{K}$$\sim$14.5-16.5; Leggett 2000c),
we estimate the distance to LL And
to be only 30 pc. 
This is far less than the lower limit of
346 pc estimated using WD models by Szkody et al., (2000). 
To obtain a distance to LL And in agreement with the Szkody et al., 
estimate, the secondary star 
would need to have M$_K$ $\sim$10.
For EF Eri, we estimate the secondary star as $\sim$L5 and 
taking M$_{K}$=11.5 (Kirkpatrick et al., 2000) 
we find a distance of 42 pc. Distance estimates for EF Eri presented 
in the literature give
d$\ga$92 pc but are based on not detecting the 
secondary star in optical spectra
and assuming it to be M5V. Using the Beuermann et al. (2000) 
limit of a $>$M9V secondary (M$_{K}$=10.5) and our same assumptions,
the distance to EF Eri would be estimated as $\la$66pc. 
Given that the secondaries in post-period minimum CVs are
likely to be only brown dwarf-like and no models yet
exist for their detailed properties,
they may have absolute magnitudes quite different from those assigned here.

The secondary stars in LL And and EF Eri are the first
brown dwarf-like objects to be unambiguously identified in a CV 
and the first direct observational 
proof of the theoretical prediction by Howell et al., (1997, 2001) 
that a unique type of sub-stellar, brown dwarf-like 
secondary star exists in very old CVs.
These binaries are the equivalent of placing 
Jupiter (equal radius but $\sim$50 times as massive)
where the moon is and having it orbit the Earth every 80 minutes.
If many post-period minimum systems (ie., TOADs)
contain similar secondary stars and if the duty cycle of actual mass transfer
in such systems is low (as it appears to be in LL And and EF Eri), 
then the majority will 
spend time in deep quiescent states and be very difficult to discover
observationally as they provide no hint of their true nature. 
Searches may need to concentrate on finding essentially
white dwarf - red (brown) dwarf pairs. Perhaps detailed observations of 
currently known short period WD+RD or WD+BD
pairs may reveal that some are, in fact, post-period minimum CVs
in deep quiescence.

\acknowledgments

The United Kingdom Infrared Telescope is operated by the Joint Astronomy Centre
on behalf of the U.K. Particle
Physics and Astronomy Research Council. The authors thank the staff 
at UKIRT, particularly Andy Adamson, Chris Davis, John Davies, Paul Hirst, 
Sandy Leggett, and Olga Kuhn,
for their enthusiasm and support of this project and their help with the
observations. Mark Huber provided assistance
with the October observations and Paula Szkody and 
John Thorstensen communicated information
regarding the optical nature of LL And. 
We are grateful to the referee for providing a number of useful comments.
We especially appreciate the help given by 
Sandy Leggett who has provided a wealth of information on 
IR spectra of low mass red
objects and also digital spectra for comparison.
SBH acknowledges partial support of this research from NSF grant AST 98-19770
and NASA Theory grant NAG5-8500.

%
\pagebreak

\newpage
\begin{figure}
\begin{center}
\psfig{file=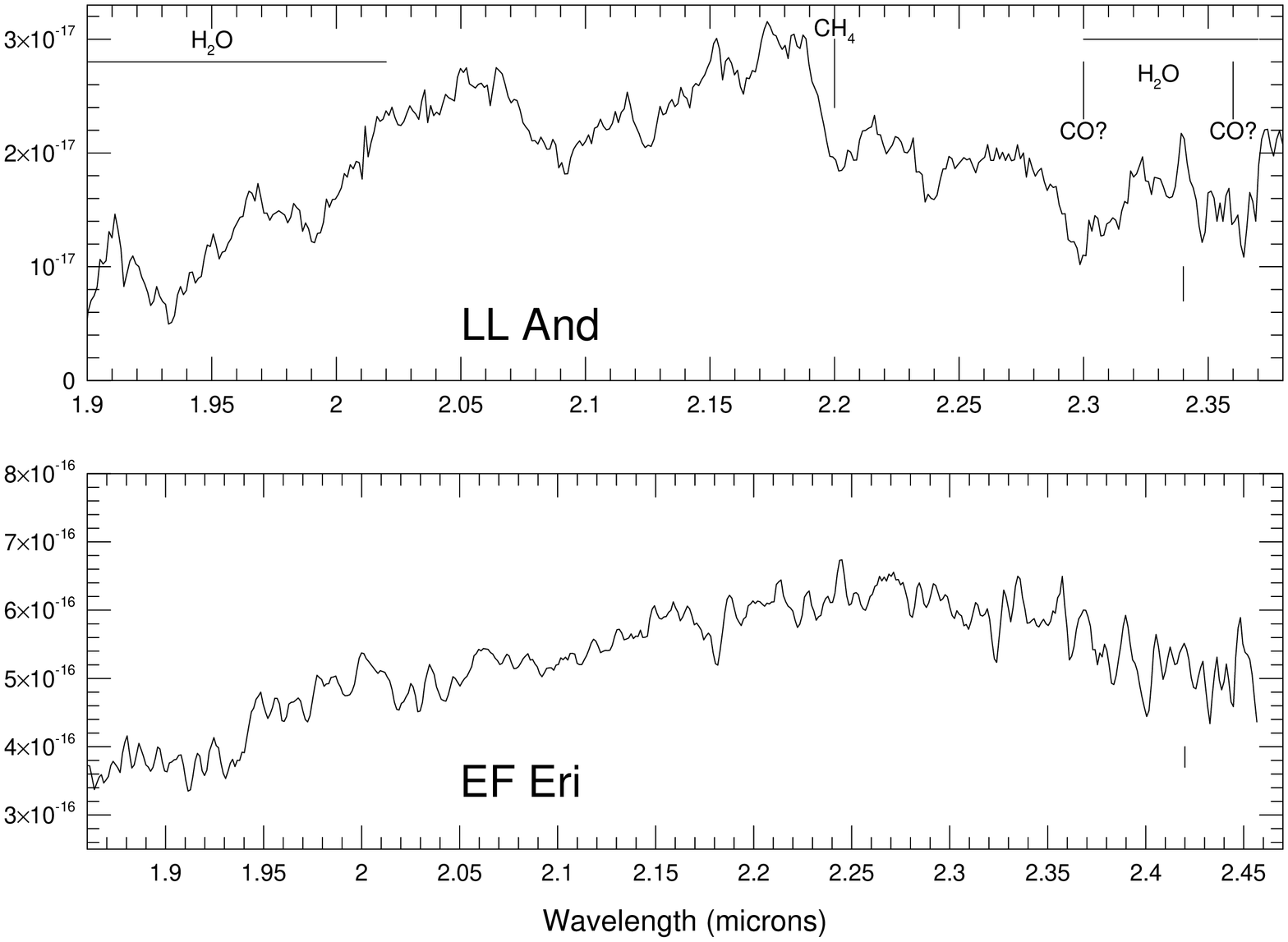,width=7.0in,angle=90}
\caption{K band spectra of LL And and EF Eri. In LL And, 
note the presence of broad 
water absorption and the
CH$_4$ absorption edge at 2.2 microns. EF Eri also shows water absorption and may
have CO bandheads. The small vertical lines in each panel are 1$\sigma$ error bars
and the flux in both panels is in units of W/m$^2$/$\mu$m.
}
\end{center}
\end{figure}

\newpage
\begin{figure}
\begin{center}
\psfig{file=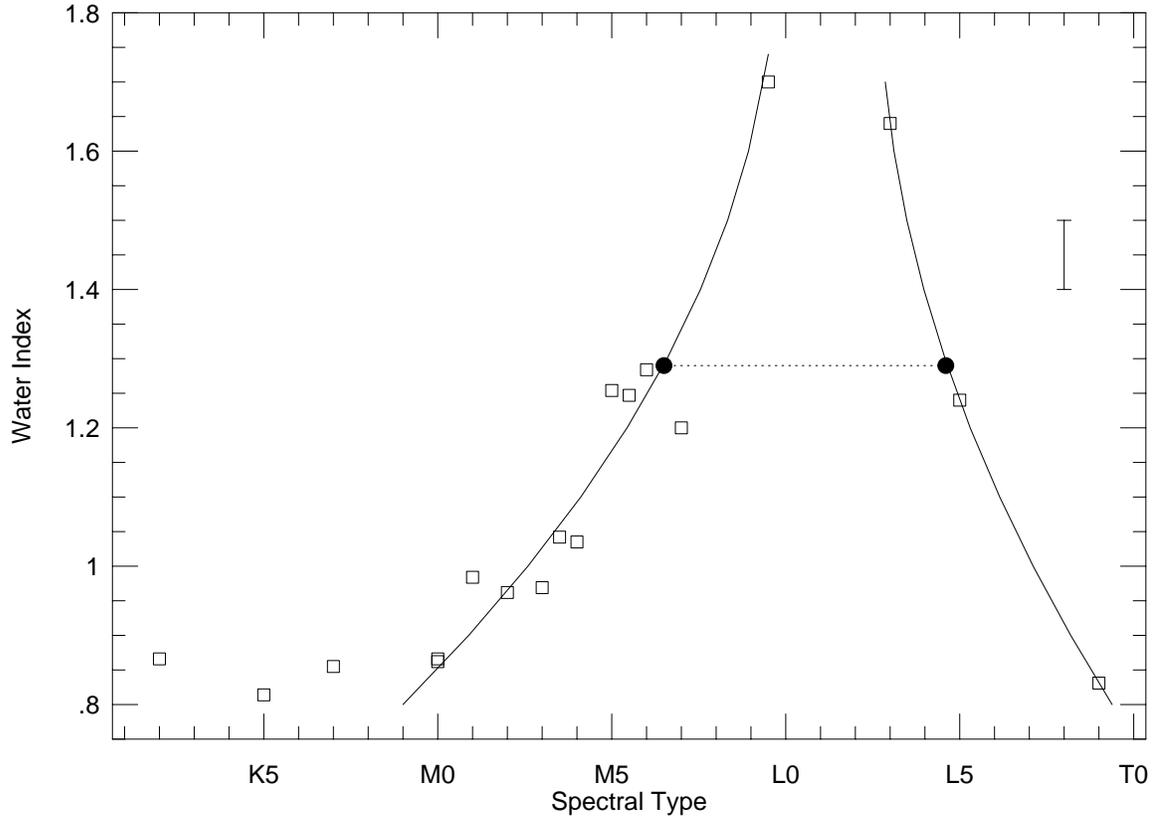,width=7.0in,angle=90}
\caption{Relationship of the water band spectral index, D$_{H_2O}$, vs. spectral
type. Open squares are measured K through L stars taken from Dhillon et al. (2000)
and Leggett et al. (2000b, 2001). The filled circles are the two possible locations
for EF Eri. The lines are simple polynomial fits to the M \& L stars
respectively. An estimate of the uncertainty in D$_{H_2O}$, due to measurement
error and/or actual spectral differences, 
is shown in the figure and was determined from the intrinsic
scatter seen in the M stars.
}
\end{center}
\end{figure}

\end{document}